\documentclass[seceq,supplement]{ptptex}
\usepackage{graphicx}
\usepackage{wrapft}

\title{
Duality symmetries in driven one-dimensional hopping  models
}

\author{
Peter Sollich$^1$ and Robert L. Jack$^2$
}

\inst{
$^1$ King's College London, Department of Mathematics,
London WC2R 2LS, UK. 
\\
$^2$ Department of Physics, University of Bath,
Bath BA2 7AY, UK.
}

\abst{
We consider some duality relations for models of non-interacting
particles hopping on disordered one-dimensional chains.  In particular,
we discuss symmetries of bulk-driven barrier and trap models,
and relations between boundary-driven and equilibrium models with
related energy landscapes.  We discuss the relationships between
these duality relations and similar results for interacting many-body
systems.
}

\begin{document}
\maketitle

\newcommand{\ppt}{\frac{\partial}{\partial t}}
\newcommand{\hf}{\frac12}
\newcommand{\hfs}{{\textstyle{\frac12}}}
\newcommand{\Bigs}{}
\newcommand{\ee}{\mathrm{e}}
\newcommand{\Sbar}{\overline{S}}
\newcommand{\Wint}{W^{(1)}}
\newcommand{\Whf}{W^{(1/2)}}
\newcommand{\Gbar}{\overline{G}}
\newcommand{\Gint}{G^{(1)}}
\newcommand{\Gbarint}{\overline{G}^{(1)}}
\newcommand{\Ghf}{G^{(1/2)}}
\newcommand{\Gbarhf}{\overline{G}^{(1/2)}}
\newcommand{\mub}{\mu_\mathrm{B}}
\newcommand{\mut}{\mu_\mathrm{T}}

\newcommand{\hEone}{\hat{E}^{(1)}}
\newcommand{\hEhf}{\hat{E}^{(1/2)}}

\section{Introduction and general hopping model}
\label{sec:model}

Among the simplest models of non-equilibrium statistical mechanics
are one-dimensional transport models.  For example, one may
consider models of particles hopping on a chain, where a current
is forced through a system either by coupling to reservoirs
at the boundaries, or by forces acting in the bulk.  Despite their
simplicity, these models exhibit rich behaviour, and are
the subject of ongoing studies~\cite{Derrida-review}.

Here, we are interested in duality relations between pairs of 
one-dimensional models.  The properties of the related models may be quite
different: for example, one may sometimes relate
non-equilibrium models to equilibrium ones~\cite{TKL-07,TKL-08,Kurchan-math,ILW-09}
or one may find mappings between models
with different realisations of disordered rates~\cite{TKL-08,schutz,JS-dual,JS-jstat}.
Our recent work has focussed on propagation of single 
(or non-interacting) particles in one dimension~\cite{JS-dual,JS-jstat},
motivated originally by properties of glassy model systems~\cite{BBL}.  
However, such models have a broad range of applications, and have been the
subject of many analytic studies~\cite{1d-general}.  The purpose
of this article is to comment on some relations between the simple
mappings that we have found and mappings in interacting (many-body) 
systems.  
In particular, a relation based on an inversion of the energy landscape
occurs in several many-body systems~\cite{schutz,TKL-08} a 
well as in our analysis~\cite{JS-jstat}.  
We aim to elucidate the origins of these mappings, particularly
a duality between sites and bonds of 1d chains: this is facilitated
by studies of simple models for which the `heavy machinery' of many-body theory
is not required.

After reviewing our previous work, which concentrated on systems
at equilibrium, we discuss a set of disordered models where  a uniform
driving force acts in the bulk.  We show how symmetries of these
models result in a factorisation of their master operators
that resemble the supersymmetric form used for time-reversible systems.
We discuss the reasons for this, despite the breaking of time-reversibility
by the driving force.  Then,
discuss duality relations between boundary-driven systems and
systems with conserved particles.  These results are related
to recent works by Tailleur, Kurchan and Lecomte~\cite{TKL-07,TKL-08}.

We first define a disordered one-dimensional hopping model for
non-interacting particles by specifying
rates for hops from site $i$ to sites $i-1$ and $i+1$,
which we denote by $\ell_i$ and $r_i$ 
respectively.  
Let $n_i(t)$ be the density of particles on site
$i$ at time $t$, with equations of motion
\begin{equation}
\frac{\partial}{\partial t}n_i(t) = \ell_{i+1} n_{i+1}(t)
 + r_{i-1} n_{i-1}(t) - (\ell_i + r_i) n_i(t).
\label{equ:dtp}
\end{equation}
for $i=1\dots N$.  
 It remains to fix the boundary
conditions.  The simplest case is to use periodic
boundaries, identifying site $0$ with site $N$.
In that case, the equations conserve the total number
of particles, $\ppt\sum_{i=1}^N n_i(t)=0$, and the
equations of motion can be interpreted as a master
operator as in Ref.~\citen{JS-jstat}.  However, we
also consider an alternative case where we
consider $n_0$ and $n_{N+1}$ as time-independent
reservoir densities, allowing a boundary-driven
system without a conserved density~\cite{TKL-08}.

We will use an operator notation, 
exploiting the linearity of the equations of motion.
We define a state 
$|n(t)\rangle=\sum_i n_i(t)|i\rangle$, 
with a basis such that $\langle i | j \rangle=\delta_{ij}$
for integer $i$.  The equations of motion are
$
\frac{\partial}{\partial t}|n(t)\rangle = W |n(t)\rangle
$
with 
\begin{eqnarray}
W
  &=& \sum_{i=1}^{N} |i\rangle \big( 
\ell_{i+1}\langle i+1| + r_{i-1}\langle i-1| - (\ell_i+r_i)\langle i| \big)
\label{equ:Wgen}
\end{eqnarray}
(The interpretation of the states $\langle 0|$ and $\langle N+1|$ which appear
in this operator depends on the boundary conditions, as described above.)

\section{Duality relations in models with conserved density and periodic boundaries}
\label{sec:dual}

\begin{figure}
\begin{center}
\includegraphics[width=6cm]{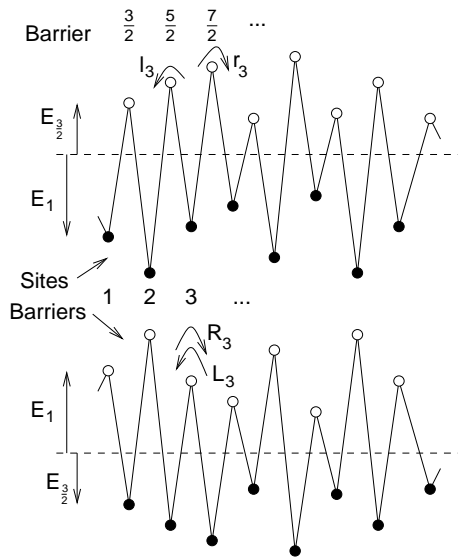}
\end{center}
\caption{
(Top) Illustration of an `energy landscape', defined in terms of
site energies $E_i$ and transition
state energies $E_{i+\hf}$.  The rate for
hopping from a site is related to the sum of the
site energy (measured downwards as shown) 
and the adjacent transition state energy (measured upwards).  
(Bottom) On inversion of the potential, the transition
state energies become energies of new sites with
half-integer indices: transition rates between
these sites are controlled by the energies $E_i$
which now have an interpretation as transition
state energies. From Ref.~\citen{JS-jstat}.
} \label{fig:invert_all}
\end{figure}

In this section, we review
some previous results~\cite{JS-jstat}, restricting our analysis
to periodic chains. In this case we have, writing $W$ as $\Wint$ to
distinguish from the dual operator $\Whf$ below,
\begin{equation}
\Wint = 
    \sum_{i=1}^{N} 
     \big( |i+1\rangle - |i\rangle \big) \big( r_i\langle i| - 
 \ell_{i+1}\langle i+1| \big).
\end{equation}
from which conservation of particles is apparent as
$\sum_i \langle i| \Wint=0$.  This allows us to intepret $\Wint$
as a master equation for stochastic motion of a single particle.

This hopping model is dual to a second
model which is of the same form, but with 
hopping on sites with half-integer indices.  That is, 
one has densities $n_{\hf}(t), n_{\frac32}(t), \dots$,
from which we construct a state 
$|\tilde n(t)\rangle=\sum_i n_{i-\hf}(t) |i-\hfs\rangle$
whose evolution is given by a master operator
\begin{equation}
\Whf = \sum_{i=1}^N  
\left( \Bigs|i+\hfs\Bigs\rangle - \Bigs|i-\hfs\Bigs\rangle \right) 
 \left( R_i\Bigs\langle i-\hfs\Bigs| - L_i\Bigs\langle i+\hfs\Bigs| \right).
\label{equ:Wb}
\end{equation}
Since the particles reside on half-integer sites, we associate
integer indices with barriers between the sites.  Thus, $L_i$
and $R_i$ are rates for motion to left and right across
the $i$th barrier.  Then, $\Wint$ is dual to $\Whf$ if we
take
\begin{equation}
L_i = r_i, \qquad R_i=\ell_i
\label{equ:lrLR}
\end{equation}
whose physical interpretation as an inversion of the energy landscape
will be discussed below (see also Fig.~\ref{fig:invert_all}).

To reveal the duality between the models, we write
\begin{eqnarray}
D &=& \sum_{i=1}^N \big(|i\rangle - |i+1\rangle \big) \langle i+\hfs|,
\nonumber \\
J &=& \sum_{i=1}^N |i+\hfs \rangle \big( r_i \langle i | - 
 \ell_{i+1} \langle i+1| \big),
\end{eqnarray}
so that 
\begin{equation}
\Wint = -DJ, \qquad \Whf = -(JD)^\dag.
\label{equ:wint_dj}
\end{equation}
This factorisation means that $\Wint$ and $\Whf$ have
the same eigenspectra.  
For example, if $\langle \psi|$ is a left eigenvector
of $\Wint$ with eigenvalue $\lambda\neq0$ then 
$D^\dag|\psi\rangle$ is a right eigenvector of
$\Whf$ with the same eigenvalue.  To express this duality
in a more standard form, we could write 
\begin{equation}
\Wint D = - DJD = D (\Whf)^\dag.
\label{equ:djd}
\end{equation}

Despite the simple form of~(\ref{equ:lrLR}), we emphasise that this
duality relates distinct pairs of hopping models.  To illustrate this,
we parameterise the rates in a region of the chain as
\begin{equation}
\ell_i = \ee^{-E_i-E_{i-\hf}}, \qquad
r_i = \ee^{-E_i-E_{i+\hf}}
\label{equ:lrE}
\end{equation}
where we interpret the $E_i$ and $E_{i+\hf}$ as site
and transition state energies.  The duality 
condition~(\ref{equ:lrLR}) can then be interpreted as 
a swap of transition state and site energies, or equivalently
as an inversion of the energy landscape, as in Fig.~\ref{fig:invert_all}.
Our sign convention is that the $E_i$ are measured downwards from 
an arbitrary baseline, so that site and transition energies are dual
to each other.  We note that a parameterisation of the rates in the form
(\ref{equ:lrE}) is always possible on any subsection of the chain, but applying
it to the whole set of rates requires additionally
a global constraint of detailed balance: $\prod_i r_i=\prod_i \ell_i$.

In addition, one may relate the
propagators within the two models.  Interpreting the equations
of motion (\ref{equ:dtp}) as a master equation for a single
particle we identify the propagator $\Gint_{ij}(t)=\langle
i|\ee^{\Wint t}|j\rangle$ as the
probability that the particle is on site $i$ given that
it was on site $j$ a time $t$ earlier.  In the description
in terms of non-interacting particles, the steady state
two-point connected correlation function is simply 
\begin{equation}
\langle n_i(t)
n_j(0) \rangle_\mathrm{ss} - \langle n_i \rangle_\mathrm{ss} \langle n_j\rangle_\mathrm{ss}
 = G_{ij}(t) \langle n_j \rangle_\mathrm{ss} 
\end{equation}
where we use the label `ss' to
indicate a steady state average.
Starting from (\ref{equ:djd}),
we consider the matrix elements $\langle i | \ee^{\Wint t} D | j -\hf\rangle
=\langle i | D \ee^{(\Whf)^\dag t} | j-\hf \rangle$,
arriving at
\begin{equation}
\Gint_{ij}(t) - \Gint_{i,j-1}(t) = \Ghf_{j-\hf,i-\hf} - \Ghf_{j-\hf,i+\hf}
\label{equ:prop_dual}
\end{equation}
where $\Ghf_{i+\hf,j+\hf}(t)=\langle i+\hfs|\ee^{\Whf t}|j+\hfs\rangle$
is the propagator in the dual model.  We emphasise that
such results have application beyond single-particle models: for
example, the same relation applies to models of diffusing
and annihilating defects in random potentials~\cite{schutz} (with
some restrictions on boundary conditions).

Importantly, Equ.~(\ref{equ:prop_dual}) allows the propagator in the dual model to be calculated
from that of the original model, without any knowledge of the disorder.
If one then chooses the $r_i$ and $\ell_i$ from (different) distributions
that are independent under translation in space then one may prove that
the disorder-averaged propagators satisfy
\begin{equation}
\Gbarint_{ij}(t) = \Gbarhf_{ij}(t)
\label{equ:dis_prop}
\end{equation}
with both sides depending only on the difference $i-j$.

Returning to models with detailed balance, $\prod_i \ell_i=\prod_i r_i$,
we now define an operator $\hEone$ such that
$\hEone|i\rangle=E_i|i\rangle$ and similarly $\hEhf$
such that $\hEhf|i+\hfs\rangle
= E_{i+\hf}|i+\hfs\rangle$.  
Then, taking
\begin{eqnarray}
S 
 = -D \ee^{-\hEhf} &=& \sum_{i=1}^{N} 
\left( |i+1\rangle - |i\rangle \right) \Bigs\langle i+\hfs\Bigs|
                 \ee^{-E_{i+\hf}}
\\
\Sbar 
 = D^\dag \ee^{-\hEone} &=&\sum_{i=1}^{N} 
\left( |i+\hfs\rangle - |i-\hfs\rangle \right) \Bigs\langle i\Bigs|
                 \ee^{-E_i}
\end{eqnarray}
we arrive at a more symmetric factorisation of the master operators~\cite{JS-jstat}:
\begin{equation}
\Wint = S\Sbar, \qquad \Whf = \Sbar S.
\label{equ:wss}
\end{equation}
We previously considered consequences of this symmetric factorisation~\cite{JS-jstat}, 
and defined a renormalisation scheme
that acts symmetrically on $S$ and $\Sbar$.  For the
purposes of this paper, the key point is that detailed balance
ensures that $\Whf \ee^{-\hEhf}=\ee^{-\hEhf}
(\Whf)^\dag$ (and similarly for $\Wint$).  This additional
symmetry of the master operators allows the symmetric factorisation
of (\ref{equ:wss}), which also implies additional `duality' relations
$\Wint S = S \Whf$ and $\Sbar \Wint = \Whf \Sbar$.
This structure appears in supersymmetric
field theories~\cite{JS-jstat,tanase-nicola,zinn-justin-susy}. In supersymmetric
models the two symmetry relations are linked to time-translation
invariance and time-reversal invariance~\cite{zinn-justin-susy} (via detailed balance and the 
fluctuation-dissipation theorem.)

\section{Bulk-driven pure trap and barrier models}

We now consider disordered systems with uniform driving forces applied
throughout the system.  This clearly breaks detailed balance and
time-reversal symmetry.  It is therefore somewhat surprising that we can identify
a restricted set of such models for which a factorisation similar to (\ref{equ:wss})
is possible.  We will discuss how this factorisation reflects symmetries
of the models under a change in the direction of the bias, at fixed disorder.

In this section, we restrict the form of the disorder to
`pure trap' and `pure barrier' models.  The former are obtained by setting
$E_{i+\hf}=0$ for all $i$, but choosing the $E_i$ freely.  Then,
$\Wint$ describes a pure trap model with $r_i=\ell_i=\ee^{-E_i}$:
site $i$ represents a trap from which particles hop to left and right with
equal probability, but the overall rate depends on the trap depth $E_i$.
We add a bias to this model by taking 
\begin{equation}
r_i = x \ee^{-E_i}, \qquad \ell_i = y \ee^{-E_i}.
\label{equ:lrE_bias}
\end{equation}
We take $x+y=2$ for convenience since this factor
simply rescales time, and we assume $x>1$ for concreteness.  
Pure barrier models are dual to pure trap models, so their sites
have half-integer indices and their master operators are
of the form $\Whf$. In the absence of a bias,
all sites have the same energy $E_i=0$, but rates for hopping between sites
depend on the barrier being crossed.  In the presence of a bias,
right- and left-going rates are multiplied by $x$ and $y$ respectively.

The master operator for the trap model is constructed 
as in (\ref{equ:wint_dj}):
\begin{equation}
W^\mathrm{rT} = \Wint = -DJ_{xy}
\end{equation}
where the label `rT' indicates that we consider a trap model biased to the right, and we
indicate explicitly that the current operator $J$ depends on $x$ and $y$ (as well as on
the $E_i$).
One can then construct a master operator,
$W^\mathrm{\ell B} = -(J_{xy}D)^\dag$ that is dual to $W^\mathrm{rT}$, in
accordance with~(\ref{equ:wint_dj}). It has barrier-crossing rates
$R_i=\ell_i=y \ee^{-E_i}$ and $L_i=r_i=x \ee^{-E_i}$ and is therefore
biased to the left, justifying our notation $W^\mathrm{\ell B}$.

As noted above, these biased models do not respect detailed balance, so they may not
be factorised as in~(\ref{equ:wss}).  
However, let $W^\mathrm{\ell T}$ be the master operator for a trap model with a bias to the left
(that is, $W^\mathrm{\ell T} = -D J_{yx}$).  It may then be verified that
\begin{equation}
W^\mathrm{\ell T} = S'_{yx} \Sbar,\qquad W^\mathrm{\ell B} = \Sbar S'_{yx} 
\label{equ:wbt}
\end{equation}
where $\Sbar$ was defined above, and 
\begin{eqnarray}
S'_{yx}
 &=& \sum_{i=1}^{N} \left( y|i+1\rangle - x|i\rangle \right) \Bigs\langle i+\hfs\Bigs|
\end{eqnarray}
Thus, $W^\mathrm{\ell B}$ has two alternative factorisations, as $-(J_{xy}D)^\dag$ from (\ref{equ:wint_dj})
and as $\Sbar S'_{yx}$.  These relations show that it is dual both to $W^\mathrm{rT}$ and $W^\mathrm{\ell T}$, and
all three operators share the same eigenspectrum. By eliminating
$W^\mathrm{\ell B}$, one can then also write a simple duality between
trap models with opposite biases,
\begin{equation}
W^\mathrm{\ell T} \ee^{\hEone} = \ee^{\hEone} (W^\mathrm{rT})^\dag
\label{equ:bias_db}
\end{equation}
This reflects a physical relation between time-reversed trajectories:
a trajectory $\pi(t)$ in the right-biased trap
model and the time-reversed trajectory $\bar{\pi}(t)$ in the left-biased
model have the same probabilities in the steady state. Note here that
both models have the same steady state densities as an unbiased trap model,
$\langle n_i \rangle_\mathrm{ss} \propto \ee^{E_i}$.

In an analogous manner we can eliminate $W^\mathrm{\ell T}$ from the duality
relations that link it to $W^\mathrm{rB}$ and $W^\mathrm{\ell B}$ to
get a duality relation between left- and right-biased barrier models:
\begin{equation}
W^\mathrm{\ell B} \Sbar D = \Sbar D (W^\mathrm{rB})^\dag
\end{equation}
where $\Sbar D = D^\dag \ee^{-\hEone} D$.
This equation also expresses a relation between propagation in the models,
under inversion both of the bias and of the direction of time. However, we did
not find any simple physical interpretaton of this relation.

\begin{figure}[t]
\begin{center}
\includegraphics[width=12cm]{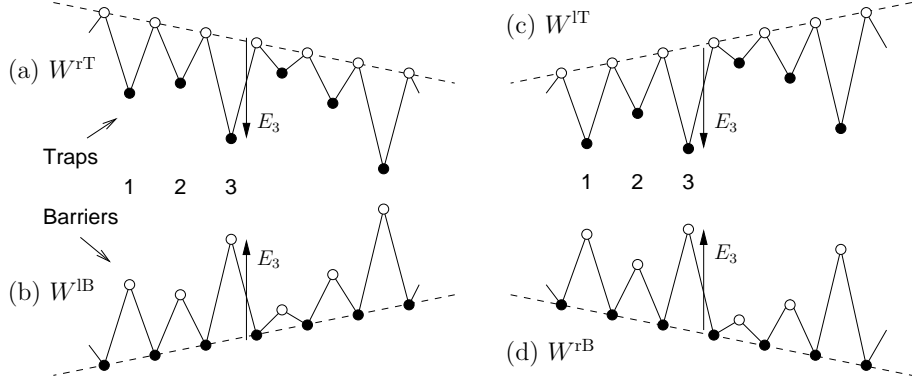}
\end{center}
\caption{
Illustration of (local) energy landscape for biased 
trap and barrier models, following Fig.~\ref{fig:invert_all}.  A landscape inversion
as in Fig.~\ref{fig:invert_all} relates (a), a trap model biased to
the right, to (b), a barrier model biased to the left.
However, Equ.~(\ref{equ:bias_db}) means that
(a) is also dual to (c), a trap model biased to the left.  Finally, inverting
the landscape (c) leads to (d), a barrier  model biased to the right.
Thus all four landscapes represent models with the same spectrum,
as described in the text.  We emphasise that the pairs (a,c) and (b,d)
are related by an inversion of the bias, but with the disordered rates $E_i$
remaining invariant.  The energy $E_3$ is identified for all four landscapes.
} \label{fig:invert_bias}
\end{figure}

To understand the relationships between the four biased trap and
barrier models, it is helpful to think in terms of tilted energy landscapes as
shown in Fig.~\ref{fig:invert_bias}. Of course the bulk bias on an entire
periodic chain cannot be represented in this way, but if we restrict
attention to propagation over finite times $t$ and therefore finite
regions of the chain, this is not an issue. For example,
to parameterise the rates~(\ref{equ:lrE_bias}) of $W^\mathrm{rT}$ in terms of
site and barrier energies as in~(\ref{equ:lrE}), one may define
$E_j' = E_j + jF$ and $E_{j+\hf}' = -(j+\hf)F - \delta$ where $F=\ln(x/y)$ 
is the driving force (in units where the temperature and lattice spacing are unity) 
and $\delta=\hf\ln(x/y)$ is an offset to the transition
state energies that simply acts to rescale the time in accordance with $x+y=2$.  The situation
is illustrated in Fig.~\ref{fig:invert_bias}(a). The factorization
$W^\mathrm{rT} = -DJ_{xy}$ leads to the dual model $W^\mathrm{\ell B}
= -(J_{xy}D)^\dag$, a pure barrier model biased to the
left as shown in Fig.~\ref{fig:invert_bias}(b). This model can
alternatively be factorized as $\Sbar S'_{yx}$, which has as its dual
$W^\mathrm{\ell T} = S'_{yx}\Sbar$, a trap model biased to the left (Fig.~\ref{fig:invert_bias}(c)).

These various symmetries also result in relations between propagators
in the various models.  If the propagator for one model (say the right-biased trap model)
is known then those of its three dual models may be constructed.  For example,
(\ref{equ:prop_dual}) reads in the notation of this section
\begin{equation}
G^\mathrm{rT}_{ij}(t) - G^\mathrm{rT}_{i,j-1}(t) = G^\mathrm{\ell
  B}_{j-\hf,i-\hf}(t) - G^\mathrm{\ell B}_{j-\hf,i+\hf}(t)
\label{equ:prop_dj_bias}
\end{equation}
independently of both bias and disorder; the labels on the propagators are the same as those on 
the associated operators $W$.
However, noting that $W^\mathrm{rT} S'_{xy}=S'_{xy} W^\mathrm{rB}$ one may
derive by a similar method
\begin{equation}
x G^\mathrm{rT}_{i,j+1}(t) - y G^\mathrm{rT}_{i,j}(t)
 = x G^\mathrm{rB}_{i-\hf,j+\hf}(t) - y G^\mathrm{rB}_{i+\hf,j+\hf}(t).
\label{equ:xy_dual_prop}
\end{equation}
which is again independent of disorder but now depends on the bias.
One may also use $\Sbar W^\mathrm{rT} = W^\mathrm{rB} \Sbar$ to arrive at
a relation that depends on disorder but not on the bias.  However, this 
last relation
reduces to (\ref{equ:prop_dj_bias}) if one notes simply that
\begin{equation}
G^\mathrm{\ell T}_{ij}(t) \ee^{E_j} = G^\mathrm{rT}_{ji}(t) \ee^{E_i}
\end{equation}
which follows from (\ref{equ:bias_db}). The latter relation between
propagation in left- and right-biased trap models can also be written
in a disorder-independent manner: for finite $t$ and hence $i-j$, the
propagators obey detailed balance with respect to the effective energy
landscape defined above, i.e.\ $G^\mathrm{rT}_{ji} \ee^{E'_i} =
G^\mathrm{rT}_{ij} \ee^{E'_j}$. This transforms our last relation into
\begin{equation}
G^\mathrm{\ell T}_{ij}(t) =  \left(\frac{y}{x}\right)^{i-j} G^\mathrm{rT}_{ij}(t)
\end{equation}
which holds independently of the disorder as anticipated.

Finally, using (\ref{equ:dis_prop}) and taking the disorder
average of (\ref{equ:xy_dual_prop}),
it may readily be shown that the disorder-averaged diffusion fronts in the 
various models are related as
\begin{equation}
\overline{G}^\mathrm{rT}_{i-j}(t) = 
\overline{G}^\mathrm{rB}_{i-j}(t) = 
\overline{G}^\mathrm{\ell T}_{j-i}(t) = 
\overline{G}^\mathrm{\ell B}_{j-i}(t)  
\end{equation} 

Thus, in this section and the preceding one, we have shown how
symmetries under time-reversal (detailed balance) or bias-inversion
may be combined with the general relation (\ref{equ:wint_dj}), leading to symmetric 
factorisations such as (\ref{equ:wss}) and (\ref{equ:wbt}).  These relations
allow the propagators of the models to be related to one another. They
also reveal an intuitively reasonable link between left- and
right-biased trap models, and a somewhat more surprising relation between left- and right-biased
barrier models.

\section{Boundary-driven case}

\newcommand{\Wd}{W^\mathrm{d}}
\newcommand{\Wr}{W^\mathrm{r}}
\newcommand{\Wa}{W^\mathrm{a}}
\newcommand{\Sd}{S^\mathrm{d}}
\newcommand{\Sdbar}{\Sbar^\mathrm{d}} 

We now turn to the boundary-driven hopping models described
in Sec.~\ref{sec:model}.  That is, the equations of
motion are given by (\ref{equ:dtp}) for $i=1\dots N$ but
we introduce two new time-independent reservoir densities,
$n_0$ and $n_{N+1}$.  
The model depends on the rates
$r_0,\ell_1,\dots,r_N,\ell_{N+1}$ which may always be parameterised
in terms of site and transition state energies as in~(\ref{equ:lrE}).  
The dependence on the reservoir densities appears
only through the combination $n_0 r_0$ and $n_{N+1} \ell_{N+1}$,
or equivalently, through $n_0 \ee^{-E_0}$ and $n_{N+1} \ee^{-E_{N+1}}$.
Since the particles are non-interacting, if we increase both
$n_0$ and $n_{N+1}$ by a multiplicative factor this simply
increases the total particle number by the same factor.  Further,
if the two reservoirs are at different chemical potentials,
$n_0 \ee^{-E_0} - n_{N+1} \ee^{-E_{N+1}} \neq 0$, there is a current
proportional to this difference.

Writing $|n(t)\rangle = \sum_{i=0}^{N+1} n_i(t) |i\rangle$
as above (but now with reservoir sites included), the equation
of motion takes the form
$
\frac{\partial}{\partial t}|P(t)\rangle = \Wd|P(t)\rangle
$
where $\Wd$ is of the form given in (\ref{equ:Wgen}). Since the 
sum in (\ref{equ:Wgen}) runs only from $1$ to $N$, we have
$\langle 0|\Wd = \langle N+1|\Wd=0$, consistent with time-independent
reservoir densities.  
It may be verified that the total
density in the model is no longer conserved, so that $\Wd$ is not
a stochastic operator: $\sum_i \langle i | \Wd \neq 0$.  

\subsection{Duality relations}

Recently, Tailleur, Kurchan and Lecomte~\cite{TKL-07,TKL-08} 
considered boundary-driven models
of interacting particles, and showed that they are dual to
`equilibrium' models that respect detailed balance.  They
further showed that this duality can be demonstrated for 
disordered non-interacting
systems, in a many-body representation.  Here, we use the methods
of the previous sections to analyse the non-interacting case.  Since
we use a basis of one-particle densities instead of a many-body basis,
our analysis is simpler.

\begin{figure}
\begin{center}
\includegraphics[width=6.6cm]{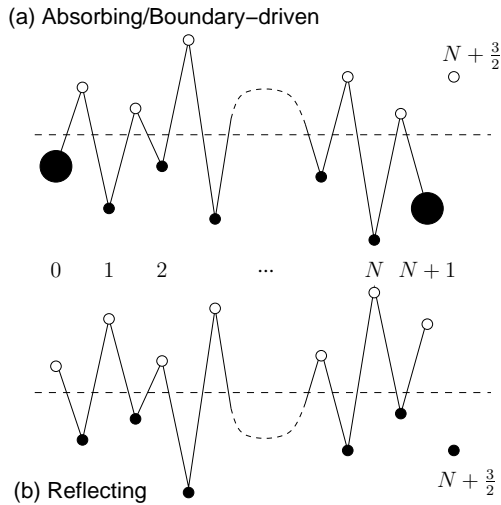}
\end{center}
\caption{Illustration of energy landscape for boundary-driven models
and models with reflecting and absorbing boundaries.  We show
only regions near the ends of the chain.  (a) The boundary-driven and absorbing
models have the same disorder: the large circles on sites $0$ and $N+1$
represent reservoirs for the driven case (operator $\Wd$) and absorbing
sites for the absorbing boundaries (operators $\Wa$).  There is an
extra transition state on site $N+\frac32$ to facilitate comparison of dual operators.
(b) The model with reflecting boundaries (operator $\Wr$).  The landscape
is obtained by inverting that for the driven/absorbing systems, following
the procedure of Fig.~\ref{fig:invert_all}.  There is an isolated
site $N+\frac32$, with no hopping in or out.  This ensures that all operators
act in spaces of dimension $N+2$. 
} \label{fig:invert_dra}
\end{figure}

We begin by factorising $\Wd$ as
\begin{equation}
\Wd = \Lambda \Sd \Sdbar
\end{equation}
with
\begin{eqnarray}
\Sd&=&\sum_{i=0}^N (|i+1\rangle - |i\rangle) 
 \langle i+\hfs| \ee^{-E_{i+\hf}},
\\
\Sdbar&=&\sum_{i=0}^{N+1} |i-\hfs\rangle ( \langle i-1| \ee^{-E_{i-1}} - \langle i|
\ee^{-E_{i}} ),
\\
\Lambda&=&\sum_{i=1}^N |i\rangle \langle i| = 1 - |0\rangle \langle 0| - |N+1\rangle \langle N+1|.
\end{eqnarray}
In $\Sdbar$, we identify site $-1$ with site $N+1$: we imagine a periodic chain with the
reservoirs sites being adjacent,
but we forbid direct hopping between the reservoirs by 
formally setting $\ee^{-E_{-\frac12}}=0$.  This makes the duality relations 
simpler since there are then equal numbers of sites $i$ and transition
states $i+\hf$.
See Fig.~\ref{fig:invert_dra}.

Then, we define two new operators
\begin{equation}
\Wr = \Sdbar \Lambda \Sd, \qquad \Wa = \Sd \Sdbar \Lambda
\end{equation}
These operators describe hopping for conserved particles, with
energy landscapes illustrated in Fig.~\ref{fig:invert_dra}.  They all have equal eigenspectra:
if $|\psi\rangle$ is a right eigenvector of $\Wd$ with
eigenvalue $\lambda\neq0$ then $\Sdbar|\psi\rangle$ is a right eigenvector of $\Wr$
and $\Sd\Sdbar|\psi\rangle$ is a right eigenvector of $\Wa$, all with
the same eigenvector.
These duality relations lead to 
the same conclusions as the many-body analysis of Ref.~\citen{TKL-08} (for the non-interacting
disordered case).  Indeed,
the many-body duality transformation considered there can be
written compactly in terms of the matrix elements of $\Sd$ and $\Sdbar$~\cite{unpub}.
Whether any similar transformation may be exploited in the systems of interacting 
particles considered in Refs.~\citen{TKL-07,TKL-08} remains a question for future work.

\subsection{Consequences of the duality symmetry}

\newcommand{\Gr}{G^\mathrm{r}}
\newcommand{\Ga}{G^\mathrm{a}}
\newcommand{\Gd}{G^\mathrm{d}}

The duality symmetry allows relations between propagators in the models
to be determined.  In particular, for the steady state in the 
boundary-driven model, we have $\langle n_i(t) n_j(0) \rangle_\mathrm{ss} - 
\langle n_i \rangle_\mathrm{ss} \langle n_j \rangle_\mathrm{ss}
 = \Gd_{ij}(t) \langle n_j \rangle_\mathrm{ss}$ with 
$\Gd_{ij}(t) = \langle i | \ee^{\Wd t} | j\rangle$.  From the relation
$\Lambda \Wa = \Wd \Lambda$, it follows that for $1\leq i,j \leq N$,
\begin{equation}
\Gd_{ij}(t) = \Ga_{ij}(t)
\label{equ:Gda}
\end{equation}
where $\Ga_{ij}(t) = \langle i | \ee^{\Wa t} | j\rangle$ is the propagator
for the model with absorbing states on sites $0$ and $N+1$.  
That is, steady state correlations in the boundary-driven steady state
of the model $\Wd$ are equal to propagation probabilities between 
sites of a model with absorbing boundaries.  Similar results
for interacting models can also be derived~\cite{Kurchan-math}.  
Physically, one notes that once a particle visits a reservoir,
all correlations associated with it are lost, so the non-trivial
correlations in the boundary-driven and absorbing models are the same.

Further, the duality between models with reflecting and absorbing boundaries
can be obtained by a factorisation, either as~(\ref{equ:wint_dj})
or as~(\ref{equ:wss}).  In the notation of this section, (\ref{equ:prop_dual})
reads:
\begin{eqnarray}
\Gr_{i-\hf,j-\hf}(t) - \Gr_{i-\hf,j+\hf}(t)
 &=& \Ga_{ji}(t) - \Ga_{j,i-1}(t) 
\nonumber \\
 &=& \Gd_{ji}(t) - \Gd_{j,i-1}(t) 
\label{equ:gra}
\end{eqnarray}
where we used (\ref{equ:Gda}) in the second equality and assumed
$1\leq i,j\leq N$. 
Remarkably, this allows the
two-point correlations of boundary-driven models to be constructed from the
equilibrium correlations of a dual model ($\Wr$) with an inverted energy landscape
but no net current.

\section{Conclusion}

We have discussed duality symmetries of non-interacting hopping models that are
revealed by factorising their equations of motion.  These factorisations indicate
symmetries of the models.  For example, if particles are conserved, master
operators may be written as $\Wint=-DJ$, allowing a dual model $\Whf=-D^\dag J^\dag$
to be constructed as in~(\ref{equ:wint_dj}).  If the models have additional
symmetry, these may be revealed through alternative factorisations such as
$\Wint=S\Sbar$ for models obeying detailed balance, or $W^\mathrm{rT}=S'_{xy}\Sbar$
for driven trap models, leading to new dual
models $\Whf=\Sbar S$ or $W^\mathrm{rB}=\Sbar S'_{xy}$.  
These factorisations and duality relations allow relationships
between propagators within the various models to be derived.  

For boundary-driven models without conserved currents, a factorisation
of the master operator exists which reveals its duality with a model where
particles are conserved and boundary conditions are reflecting, which we intepret
as an `equilibrium' model.  This then also allows a further relation to be derived between
the boundary-driven model and one with absorbing boundaries.  

Several of these relations are known in models of interacting particles~\cite{TKL-07,TKL-08,
ILW-09,Kurchan-math,schutz}, although derivations in such models
require considerably more detailed calculations.  Whether the methods presented here
can be used to interpret and generalise such relations remains a topic for future
study.

\section*{Acknowledgements}

We thank the organisers of the workshop ``Frontiers in non-equilibrium
physics'' at the Yukawa Institute in Kyoto, where
many of these results were obtained.  We thank Jorge Kurchan and Fred
van Wijland for helpful discussions, and Gunter Sch\"utz for pointing
out the links to Ref.~\citen{schutz}.


\begin{thebibliography}{99}
\bibitem{Derrida-review}
For a review, see, for example, B.~Derrida, J. Stat. Mech (2007) P07023.

\bibitem{TKL-07}
J.~Tailleur, J. Kurchan and V.~Lecomte, Phys. Rev. Lett. {\bf99}  (2007), 150602.

\bibitem{TKL-08}
J.~Tailleur, J. Kurchan and V.~Lecomte, J. Phys. A {\bf41} (2008), 505001.

\bibitem{Kurchan-math}
C.~Giardina, J.~Kurchan, F.~Redig and K.~Vafayi, J. Stat. Phys. {\bf 135} (2009), 25.

\bibitem{ILW-09}
A.~Imparato, V.~Lecomte and F.~van Wijland, Phys. Rev. E {\bf 80} (2009), 011131.

\bibitem{schutz}
G. M. Sch\"utz, Z. Phys. B {\bf 104} (1997), 583;
G. M. Sch\"utz and K. Mussawisade, Phys. Rev. E {\bf 57} (1998), 2564.

\bibitem{JS-dual}
R.~L.~Jack and P.~Sollich, J. Phys. A {\bf41} (2008), 324001.

\bibitem{JS-jstat}
R.~L.~Jack and P.~Sollich, J. Stat. Mech (2009), P11011.

\bibitem{BBL}
E.~Bertin, J.-P.~Bouchaud and F.~Lequeux, Phys. Rev. Lett. {\bf 95} (2005), 015702;
R.~L.~Jack, P.~Sollich and P.~Mayer, Phys. Rev. E {\bf 78} (2008), 061107. 

\bibitem{1d-general}
For reviews, see
S. Alexander \emph{et al.}, Rev. Mod. Phys. {\bf53} (1981), 175;
J.-P.~Bouchaud and A.~Georges, Phys Rep {\bf195} (1990), 127;
R.~Metzler and J.~Klafter, Phys. Rep. {\bf339} (2000), 1.

\bibitem{tanase-nicola}
S.~Tanase-Nicola and J.~Kurchan, Phys. Rev. Lett. {\bf 91} (2003), 188302;
S.~Tanase-Nicola and J.~Kurchan, J. Stat. Phys. {\bf 116} (2004), 1201.

\bibitem{zinn-justin-susy}
J. Zinn-Justin, \emph{Quantum Field Theory and Critical Phenomena},
chapter 17
(Oxford University Press, Oxford, 2002)

\bibitem{unpub}
R.~L.~Jack and P.~Sollich, unpublished.

\end{thebibliography}
\end{document}